# An Investigation of Face Mask Use with Busking Videos on YouTube during COVID-19: a Case Study in South Korea

Chen Wu[1#], Xingjie Hao[2#], Meiqi Hu[1], Chengguqiu Dai[2], Bo Du[3*], Liangpei Zhang[1]

1. State Key Laboratory of Information Engineering in Surveying, Mapping and Remote Sensing, Wuhan University

2. School of Public Health, Tongji Medical College, Huazhong University of Science and Technology

3. School of Computer Science, Wuhan University

## Abstract

Wearing face mask is an effective measure to reduce the risk of COVID-19 infections and control its transmission, thus its usage survey is important for better policy decision to mitigate the epidemic spread. Current existing worldwide surveys are mostly self-reported, whose accuracies are hard to guaranteed, and may exaggerate the percentage of face mask wearing. Therefore, we collected busking videos with a large amount on YouTube from December 2019 to December 2020, mainly from South Korea, and reported an objective investigation of face mask use in the crowds outdoor. It is found that the face mask wearing rate has an obvious positive correlation with effective reproductive number ($R_t$) in the South Korea, which indicates that the people in South Korea kept sensitive to the COVID-19 epidemic. The face mask wearing rate in South

Korea is higher than some other countries, and two rate droppings in June and September also corresponds to the temporary remission in 2020. This study shows significant potentials to utilize public big video data to make an accurate worldwide survey of face mask use with the support of deep learning technology.

**Keywords**: Face mask use, COVID-19, busking videos, deep learning.

## Introduction

The coronavirus pandemic has changed all the lives of human being across the world completely [1]. By July, 2022, more than 500 million confirmed cases of COVID-19, including 6 million deaths, are reported by WHO [2]. The usage of face mask is proved as an effective measure to reduce the risk of COVID-19 infections and control its transmission [3-7]. Wearing face mask is also widely advocated by WHO and many governments to mitigate the spread of epidemic [8-10].

The survey of wearing face mask corresponding to locations and time is essential to study whether preventive measures are working, forecast the pandemic's spread, and to assist public health agency partners [5,11-17]. There are also many survey sources about face mask wearing offered by universities, traditional media company, social network platform and data company, etc. [18-22] However, most of these surveys are based on self-report, such as surveys by Facebook [18-20] or other network platform [14,17,23], interviews [22], and telephone [24]. The accuracy of self-reported information cannot be guaranteed [15], and these biased information may exaggerate the percentage of face mask wearing [25].

Other objective survey methods include organizing data collectors to conduct direct observation and recordation of face mask use [11,26], such as the MASCUP (Mask Adherence Surveillance at College and Universities) study covering over 60 colleges and universities across the United States [27,28]. The direct observation is the most credible survey method, whereas it is so labor- and time-consuming that this survey can only be implemented in a limited range. An alternative survey is indirect observation by camera videos to count the percentage of mask wearing [29]. It provides great potentials to implement objective and large-scale surveys. However, the problem is, it can only be conducted in one city, state, or country, since no governments are willing to share their security camera records to the public.

We are inspired by the existing surveys that, can we find shared big video data during the COVID-19 epidemic all over the world, which were recorded in public places and reflected the actual mask usage of crowds? These videos should have nothing to do with the surveys or interviews for COVID-19, so that it can avoid bias to count the mask wearing. Therefore, we choose the busking video uploaded on YouTube.

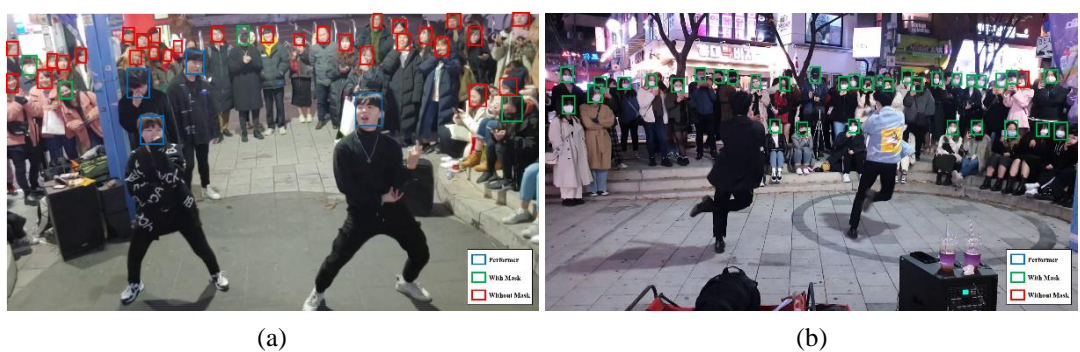

(a)                          (b)

Figure 1. Busking videos from the channel of "LUV-busking" recorded in South Korea on (a) 2020-02-08, and (b) 2020-11-02, where the face mask wearing rate in the crowds increased from 9% to 97% both in winter.

Busking indicates the street performance that performing act in the public places [30].

Since busking videos are mostly recorded in the public places with a large volume of pedestrian traffic and not special for the pandemic, they truly reflect the usage of face mask in crowds. The videos are uploaded in lots of countries all over the world and throughout the epidemic, so that we can study the tendencies with respect to time and locations. As shown in Figure 1, the busking videos recorded before and after the COVID-19 epidemic reflect the different percentages of mask wearing. The current development of deep learning technology can also help us to make the survey automatically dealing with such a big volume of busking video data [31,32].

Therefore, in this study, we collected thousands busking videos from various YouTube channels from December, 2019 to December, 2020. A face and mask detection deep learning model provided by PaddleHub [33] was used to count for the crowds and mask usage. Since only the busking videos in South Korea have an abundant quantity for statistical analysis, this paper mainly focusses on the face masking wearing rate corresponding to the epidemic situation. These busking videos in South Korea contain 3535 videos, 297.94 hours, and 10.18 million detected faces in 220 days throughout the whole year in 2020. We hope that this study can provide a new aspect of view in the survey of face mask use in COVID-19 epidemic.

# Results

*Face Mask Wearing in South Korea*

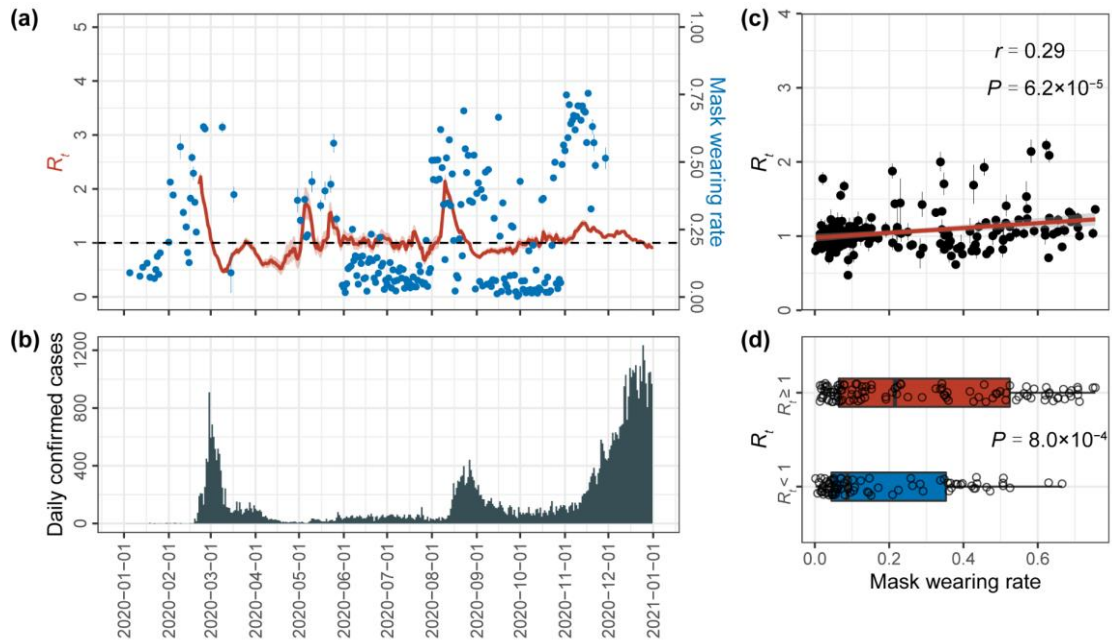

**Figure 2. The correlation of daily mask wearing rate and estimated $R_t$ in South Korea.** (a) Daily mask wearing rates and estimated $R_t$ of 2020 in South Korea. The blue points showed the daily mask wearing rates and the error bars showed the 95% confidence interval. The red line along with the shadow demonstrated the estimated $R_t$ and 95% Bayesian credible interval. $R_t$ was estimated from the day 7 (February 22, 2020) of the first epidemic peak. (b) Daily confirmed cases of 2020 in South Korea. (c) Mask wearing rates were positively correlated with $R_t$ using Spearman correlation coefficient. The error bars showed the 95% Bayesian credible interval for $R_t$. The red line was the fitted linear regression line. (d) The difference of mask wearing rates between $R_t < 1$ and $R_t \geq 1$ groups were statistically significant by Mann-Whitney U test.

The face mask wearing rates in YouTube Videos, daily effective reproductive number ($R_t$) and confirmed cases in South Korea corresponding to date are shown in Figure 2. It can be observed that, the face mask wearing rates show high positive relationship with $R_t$. The relationship is also indicated with the linear regression, with the correlation coefficient $r$ of 0.29, and the significance level $p$-value of $6.2 \times 10^{-5}$. In addition, the mask wearing rate were also higher in days with $R_t \geq 1$ compared to that with $R_t < 1$ ($p$-value = $8.0 \times 10^{-4}$).

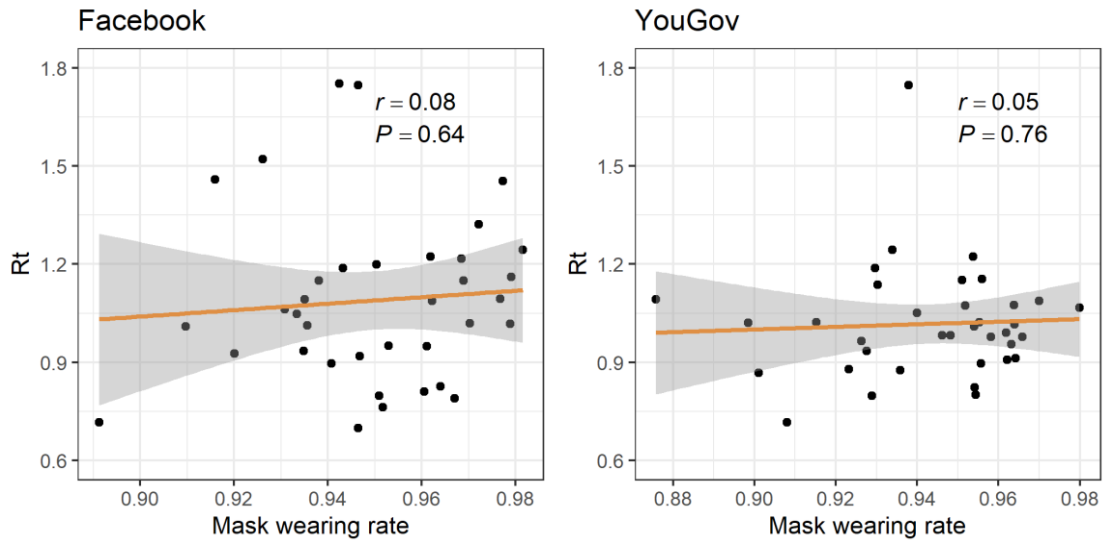

**Figure 3. The correlation of weekly reported mask wearing rate and estimated $R_t$ in South Korea.** The weekly reported mask wearing rates were from Facebook (left) and YouGov (right). The Spearman correlation test were conducted between the mask wearing rates and $R_t$. The orange line was the fitted linear regression line.

We also collected the self-reported face mask wearing rates from the widely used Facebook and YouGov, which could be exaggerated and almost all were above 90% (Figure 3). We did not observe the obvious trends of positive correlations between the mask wearing rate and $R_t$ ($p$-value > 0.5).

This result indicates that, individuals in South Korea pay close attention to the COVID-19 epidemic. When the epidemic spread faster, the face mask wearing rate increases obviously; when the mitigation of epidemic appears, person in outdoors tend to not wear face mask, even in the crowds. It can also indicate that, compared with the self-reported investigations by Facebook and YouGov, the face mask wearing rate estimated from YouTube video data shows higher precision so as to better discover this relationship.

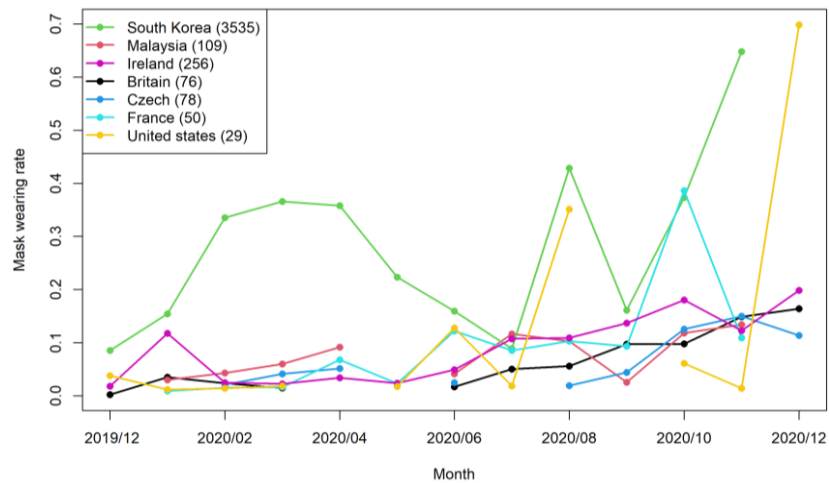

**Figure 3. The monthly statistical results of face mask wearing rate from YouTube data.** The figure shows the mean rate in the countries with more than 10 videos. The number behind the legend shows the video count of each country used in this figure.

Although the YouTube busking videos in other countries have no enough quantities for statistical analysis, we can make a quantitative comparison to partly show the difference of face mask wearing throughout the world. Figure 3 show the mask wearing rate corresponding to each month in the countries with more than 10 videos. The video count is shown in the right.

It can be seen that, compared with other countries, people in South Korea wear more masks in the whole year of 2020. From the tendency, the face mask wearing rates in these countries increase continuously. Almost 70% persons in South Korea will wear face masks even in the outdoor environments in the end of 2020, which are higher than some other countries. This result can be explained due to the stringent government response [37] and the culture of collectivism in east Asia [14].

*Limitation*

Although we have tried our best to improve our experiments, there are still some limitations. The limitations listed as follows is beneficial for understanding our study, and can also provide information for future works.

(1) Even though we have collected all available busking videos in the period on YouTube, the total count is only 4320. More than 80% collected videos were uploaded from South Korea. These videos can indicate some temporal and spatial tendencies of face mask usage. However, more videos are better when analyzing big video data to avoid some biases.

(2) Almost all busking videos were recorded outdoor. Although mask wearing is also advocated in outdoor environments [10], the demand is not as strict as that in indoor public places as shopping centers and schools [34]. But since the busking are mostly performed in crowds, face mask is also necessary to avoid exposure [10,35]. So, the investigation of face usage with busking videos is still meaningful.

(3) The mask detection on busking videos are implemented by PyramidBox-Lite deep learning model provided by PaddleHub[33]. By our observation, some detection errors will appear when the persons are far away from the camera, and the model cannot distinguish face mask with camera blur. Other factors, such as face masks with special colors or design, will also confuse the artificial intelligence model. In summary, the percentage of mask usage is more likely to be underestimated. However, since all the videos are processed with the same model, the temporal tendencies and comparison relationships will keep right.

(4) The mask usage of performers is not included in the investigation, since mask wearing may influence their performances and their faces will always appear in most frames of the videos. In order to avoid this problem, we use a python module named face_recognition from GitHub[36] and pre-selected performer face samples of each video to filter out performer faces. Since the camera positions and angles change a lot in the busking videos and the performers may show quite different poses, the filter out process will not be very accurate. However, this impact is the same for all busking videos, so the tendencies and comparisons of mask usage are still meaningful.

## Discussion

Our results indicate that the face mask usage continuously increased from December 2019 to December 2020, covering the whole period from the outbreak of COVID-19 to its rapid spread. Almost 70% persons in South Korea will wear face masks even in the outdoor environments in the end of 2020, which are higher than some other countries and can be explained by the culture of collectivism [14]. The face mask wearing rate in the South Korea shows an obvious positive correlation with $R_t$. It also indicates that persons in the South Korea kept sensitive to the COVID-19 epidemic, and wear more masks when the coronavirus spread faster.

It can be found that there are some waves in face mask usages as shown in Figure 3. Some are caused by the quality of collected data, since the counts of busking videos in some countries are not enough to balance the biases. However, we can find that in most figures, there are two troughs of waves in the June and September of 2020 (Figure 3). We suppose that the first trough in June is due to the hot weather in the summer of the Northern Hemisphere, which makes lots of persons take off their masks outdoor; the second trough in September is because in many countries, the COVID-19 epidemic seems to be mitigated [2] and many governments eased their restrictions at that time [38]. However, a worse COVID-19 outbreak in the world made persons wear mask on their faces again in the winter of 2020 [2], and the mask usages in all the countries increased in the investigations by busking videos.

Face mask usage is an effective measure to control COVID-19 transmission, and is recommended by many scientists and governments [7-10]. Surveying on personal mask usage is necessary for epidemiologic studies and policy decision [19]. However, most existing worldwide surveys are self-reported [18,20,22], which may exaggerate the mask usage [29]. Therefore, in this paper, we collected busking videos published on YouTube from December 2019 to December 2020, and make an objective investigation of face mask wearing in the crowds outdoor. The mask usage was determined by deep learning network, and the temporal/spatial mask wearing throughout the world was surveyed from the big video data.

The significance of our study lies in two aspects. First, to our best knowledge, this is the first objective investigation of worldwide face mask use covering the whole period from the beginning of COVID-19 to the end of 2020. The crowds in busking videos reflected the true situation of outdoor mask wearing.

The second significance lies in the investigation method. This study provides significant potentials to utilize shared big video data to make a worldwide survey of face mask use, which is more accurate and objective than the self-reported survey. The artificial intelligence technology supports the rapid mask identification and statistics in video data without the labor of interpreters, which make it possible to take full advantage of public videos, such as short videos in tik-tok or free webcams online if available.

## Method

*Data Collection*

In order to better investigate the usage of face mask, we collected all the videos with the keyword of "busking" and "street performance" on YouTube from December 2019 to December 2020. We only selected the videos that were really recorded on the street and have crowds or pedestrians passing by in the frames.

There are only a few videos show their record locations and time in their titles and corresponding descriptions. So, we assigned the videos with the locations of channel registrations and publish time, which are corrected with the accurate information in the titles and descriptions if they exist. It is true that in this way there will be some errors in video record time and locations, such as the authors delay to publish recorded videos, or publish old videos, or post videos recorded in other locations. However, we believe that these errors only occupy a small proportion and will not have a great influence in the statistics of big video data.

Finally, we filter out the unqualified videos, such as only the performers appear in the videos or the record time is out of range, and obtained a total number of 4320 busking videos. They are published by 32 YouTube channels and contain more than 400 hours. Their locations include 17 countries, mostly distributed in Asia, Europe, and North America. In this paper, we mainly focus on 3535 videos uploaded from South Korea, containing 297.94 hours, and 10.18 million detected faces in 220 days throughout the whole year in 2020.

*Mask Detection*

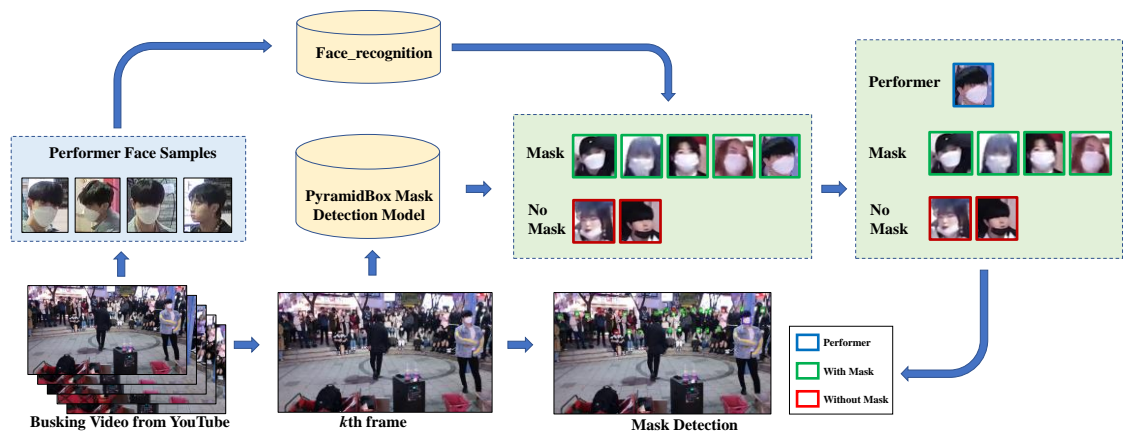

Figure 4. Flowchart of the mask detection.

Since we have more than 400 hours videos to observe, it is impossible to count for mask wearing manually. Therefore, we utilize deep learning network model to detect faces from the frames and identify mask wearing in each video. The flowchart is shown in Figure 4.

It is worth noting that the performers are special in the process of mask detection for two reasons: 1) the performers may not wear mask due to their performance requirement; and 2) the performers will appear in almost every second of the videos since they are the leading roles. So, it is better to filter out the performers and only consider the crowds or passerby to reflect the real situation.

The whole procedure of mask detection with busking videos is as follows:

Firstly, for each video, we split one frame in every one second as the process unit. A deep learning model named PyramidBox-Lite provided by PaddleHub[33] was used to detect faces and identify mask wearing.

Then, we extract performer face samples from each video. A python module named face_recognition from GitHub[36] was used to encode the faces of the performer and the detected faces by PyramidBox-Lite. The detected faces were compared with the extracted performer face samples, and the performer faces will not be included in the final statistics of face mask wearing.

Finally, all the detected faces with mask wearing labels were recorded for each video. The percentage of mask use was calculated excluding performers in each video. The tendencies of mask use are summarized by the unit of one video.

*Statistical analysis*

Mask wearing rates obtained from the YouTube videos can be an effective measurement to assess the mask usage in populations. Taking 1 day as a unit, we computed the daily mask wearing rate by dividing the total masks wearing counts for a certain day to the total faces detected on the same date.

To investigate the relationship between face mask wearing and the transmission of COVID-19, we estimated the time-varying reproduction number ($R_t$) of South Korea in 2020 using EpiEstim (v. 2.2-4) [39]. $R_t$ was estimated from the day 7 (February 22, 2020) of the first epidemic peak based on the daily confirmed cases obtained from WHO [39] and a parametric serial interval of South Korea in 2020 with the mean of 3.93 days and the standard deviation of 4.86 days [40]. Subsequently, we calculated the Spearman correlation coefficient $r$ to assess the correlation between $R_t$ and mask wearing rates. Further, we also compared the mask wearing rates between $R_t \geq 1$, which indicates the spread of the epidemic, and $R_t < 1$ using Mann-Whitney U test.

Further, we also obtained the mask wearing rates from the COVID-19 questionnaires of people in South Korea on YouGov [14] and Facebook [41]. For the questionnaires from YouGov, we computed the weekly mask wearing rates based on the question "Worn a face mask outside your home (e.g. when on public transport, going to a supermarket, going to a main road)". Five levels can be chosen, including "Always", "Frequently", "Sometimes", "Rarely", and "Not at all". To generate a binary measurement, we considered people who chose "Always" and "Frequently" as wearing a mask and those who chose the others as not wearing a mask. The week interval has been defined by YouGov and the weekly mask wearing rates were calculated thereafter. Regarding the questionnaires on Facebook, we acquired the weekly mask wearing rates of the participants from South Korea between April 26, 2020 and January 2, 2021.

All the statistical analyses were conducted in R (v. 4.0.2) software. *P* < 0.05 was considered to be statistically significant.

## Acknowledgements:

This work was supported in part by the National Natural Science Foundation of China under Grant 61971317 and 61822113.

## Competing interests:

The authors declare no competing interests.

## Author contributions:

C. W. designed the research framework, developed the algorithm, analyzed the results, and wrote/edited the paper. X. H. implemented the statistical analysis, analyzed the results, and wrote/edited the paper. C. W. and X. H. contributed equally to this work. M. H. cleaned the video data, labeled the performer samples, and analyzed the results. C. D. analyzed the results. B. D. and L. Z. contributed to the designing of research and the writing of the paper.

## Corresponding Authors

Correspondence and requests for materials should be addressed to Bo Du.

# Data Availability Statements

All the video data are available on YouTube. The code that supports the findings of this study are available on request from the corresponding author.